\documentclass[conference, twoside]{IEEEtran}
\usepackage{fancyhdr} %used for header

\usepackage{lipsum}
\usepackage[utf8]{inputenc}

\usepackage{authblk}

\usepackage[T1]{fontenc}

\usepackage{listings}
\usepackage{xcolor}

\definecolor{codegreen}{rgb}{0,0.6,0}
\definecolor{codegray}{rgb}{0.5,0.5,0.5}
\definecolor{codepurple}{rgb}{0.58,0,0.82}
\definecolor{backcolour}{rgb}{0.95,0.95,0.92}

\lstdefinestyle{mystyle}{
    backgroundcolor=\color{backcolour},   
    commentstyle=\color{codegreen},
    keywordstyle=\color{magenta},
    numberstyle=\tiny\color{codegray},
    stringstyle=\color{codepurple},
    basicstyle=\ttfamily\footnotesize,
    breakatwhitespace=false,         
    breaklines=true,                 
    captionpos=b,                    
    keepspaces=true,                 
    numbers=left,                    
    numbersep=5pt,                  
    showspaces=false,                
    showstringspaces=false,
    showtabs=false,                  
    tabsize=2
}

\usepackage[absolute,overlay]{textpos} % for absolute positioning

\usepackage{graphicx} %% for loading jpg figures
\usepackage{subfig}
\usepackage{nccmath}
\usepackage{color}
\usepackage{cite}
\usepackage{acro}
\usepackage{scalerel}
\usepackage{tikz}
\usetikzlibrary{svg.path}

\usepackage{orcidlink}

\definecolor{orcidlogocol}{HTML}{A6CE39}
\tikzset{
  orcidlogo/.pic={
    \fill[orcidlogocol] svg{M256,128c0,70.7-57.3,128-128,128C57.3,256,0,198.7,0,128C0,57.3,57.3,0,128,0C198.7,0,256,57.3,256,128z};
    \fill[white] svg{M86.3,186.2H70.9V79.1h15.4v48.4V186.2z}
                 svg{M108.9,79.1h41.6c39.6,0,57,28.3,57,53.6c0,27.5-21.5,53.6-56.8,53.6h-41.8V79.1z M124.3,172.4h24.5c34.9,0,42.9-26.5,42.9-39.7c0-21.5-13.7-39.7-43.7-39.7h-23.7V172.4z}
                 svg{M88.7,56.8c0,5.5-4.5,10.1-10.1,10.1c-5.6,0-10.1-4.6-10.1-10.1c0-5.6,4.5-10.1,10.1-10.1C84.2,46.7,88.7,51.3,88.7,56.8z};
  }
}

\newcommand\orcidicon[1]{\href{https://orcid.org/#1}{\mbox{\scalerel*{
\begin{tikzpicture}[yscale=-1,transform shape]
\pic{orcidlogo};
\end{tikzpicture}
}{|}}}}

\ifCLASSINFOpdf
\else
\fi
\DeclareAcronym{acm}{
  short = ACM ,
  long  = Association for Computing Machinery ,
  sort  = A ,
}

\usepackage{hyperref} %<--- Load after everything else
\begin{document}
 
%%%%%%%%%%%%%%%%%%%%%%%%%%%%%%%5
%Se ha comentado el NUMERO DE PAGINA
% \setcounter{page}{7}%%%% HERE SET THE PAGE NUMBER

%%%%%%%%%%%%%%%%%%%%%%%%%%%%%

\title{

IoT-Driven Smart Management in Broiler Farming: Simulation of Remote Sensing and Control Systems

%Smart Broiler Management with IoT: Simulation of Remote Sensing and Control Systems
}
%
%

%\author{Sandra Isabella Coello \orcidicon{0009-0003-2447-2598}\, V. Sanchez Padilla \orcidicon{0000-0003-3205-388X}\,,\IEEEmembership{ Member, IEEE}, Ronald Ponguillo-Intriago \orcidicon{0000-0002-7554-8367}\,% <-this % stops a space
%\thanks{Manuscript received Month Day, Year; revised Month Day, Year.}
%\thanks{Sandra Isabella Coello is with the Escuela Superior Politécnica del Litoral, Guayaquil, EC090112, Ecuador.} % (e-mail:saiscoel@espol.edu.ec)}% <-this % stops a space
%\thanks{V. Sanchez Padilla is with the Universidad ECOTEC, Km 13.5 Samborondón, Samborondón, EC092302, Ecuador.} %, and also with the Virginia Polytechnic Institute and State University, Blacksburg, VA 24060, USA (e-mail: vsanchez@vt.edu).} % <-this % stops a space
%\thanks{Ronald Ponguillo-Intriago is with PoInt, Guayaquil EC090112, Ecuador.} % (e-mail: ronald.ponguillo@gmail.com)} % <-this % stops a space}

%\author[]{N.N.}
\author[1]{Sandra Coello Suarez}
\author[2,3]{V. Sanchez Padilla\orcidlink{0000-0003-3205-388X}}
\author[4]{%\textit{Member, IEEE}
Ronald Ponguillo-Intriago\orcidlink{0000-0002-7554-8367}}
\vspace{1pt}
\author[1]{Albert Espinal\orcidlink{0000-0003-2619-2752}}
\affil[1]{Telematics Engineering Dept., Escuela Superior Politécnica del Litoral, Guayaquil, 090112, Ecuador}
\affil[2]{Dept. of Engineering Education, Virginia Tech, Blacksburg, VA 24060, USA}
\affil[3]{Faculty of Engineering, Universidad ECOTEC, Samborondón, EC092302, Ecuador}
\affil[4]{Facultad de Ciencias Matemáticas y Físicas, Universidad de Guayaquil, Guayaquil, 090313, Ecuador}
\affil[ ]{%\textit
{\{saiscoel, aespinal\}@espol.edu.ec, vsanchez@vt.edu, rponguil.point@gmail.com %ronald.ponguilloi@ug.edu.ec
}}

\maketitle

%First, add: \usepackage[absolute,overlay]{textpos} % for absolute positioning

% Absolute positioning of copyright notice
\begin{textblock*}{\textwidth}(15mm,5mm) % adjust vertical position here
\centering
\footnotesize © 2025 IEEE. This is the authors’ version of the work accepted for presentation at the 2025 IEEE Technology and Engineering Management Society Conference (TEMSCON LATAM), Cartagena, Colombia. The final version will be available in IEEE Xplore. Personal use of this material is permitted. Permission from IEEE must be obtained for all other uses.
%, including reprinting/republishing for advertising or promotional purposes, creating new collective works, for resale or redistribution to servers or lists, or reuse of any copyrighted components of this work in other works. %The final version is available at: \url{https://doi.org/DOI}

\end{textblock*}

\begin{abstract}

%\lipsum[2-3]

Parameter monitoring and control systems are crucial in the industry as they enable automation processes that improve productivity and resource optimization. These improvements also help to manage environmental factors and the complex interactions between multiple inputs and outputs required for production management. This paper proposes an automation system for broiler management based on a simulation scenario that involves sensor networks and embedded systems. The aim is to create a transmission network for monitoring and controlling broiler temperature and feeding using the Internet of Things (IoT), complemented by a dashboard and a cloud-based service database to track improvements in broiler management. We look forward this work will serve as a guide for stakeholders and entrepreneurs in the animal production industry, fostering sustainable development through simple and cost-effective automation solutions. The goal is for them to scale and integrate these recommendations into their existing operations, leading to more efficient decision-making at the management level.

\end{abstract}

% Note that keywords are not normally used for prerreview papers.

\IEEEpeerreviewmaketitle

\section{Introduction}

\IEEEPARstart{M}{any} farmers responsible for broiler breeding use manual or informal methods without automation due to a lack of proper training in technology or logistics \cite{istiak2022poultry, bhanja2021automation, Sonaiya2004}. Since broilers are a huge source of food for people, both for their meat and derivatives, they also represent a valuable income for farmers. Important parameters in broiler breeding include temperature and food supplementation by time intervals, in which control can be challenging based on available resources and assets \cite{astill2020smart, Cassuce2013}. An appropriate ambient temperature should follow the recommended range or thresholds, or broiler breeding will be directly or indirectly affected by the wrong setting of these critical parameters, resulting in animal stress \cite{vsrankova2019floor, nawab2018heat, Moura2015}. Farmers should also consider levels of food supply, for example, if it is low, it leads to a low weight group, with unsuitable productivity outcomes \cite {Cassuce2013}. Excessive eating and an excess of nutrients provided to the broiler also lead to inappropriate fattening, which could have negative consequences for broiler management \cite{kucheruk2019comparison}.

A way to address the problem described is to set up a monitoring system to visualize and control environmental temperature and food distribution with alerts \cite{10010720, 9612437, Edwan2020}. Current technology allows more options and ease of connection to the Internet to support process automation, which enhances production. The Internet of Things (IoT) includes network connectivity concepts to improve transmissions and control in home and industrial systems through protocols and technologies to enhance productivity, increase efficiency, reduce cost production and meet business demands \cite{mancheno2022development, Khujamatov2021}. Thus, we propose a wireless sensor network simulation to monitor and control broiler production through a network system that triggers warnings based on changes in ambient temperature thresholds and food supply.

The first stage of this work consists of sensors that transmit data on temperature and food load values for food storage and monitor the breeding process to optimize resources. Sensors control the threshold temperature based on a reading process in which the system verifies the temperature by comparing the values for a subsequent activation of the actuator, such as a fan or lamp, to stabilize the proposed temperature. The second stage enables us to know if food is required to supply the broiler based on the results of a load sensor that measures its weight. An Arduino transmits data to a Raspberry Pi through serial communication to save it and present it on a dashboard to allow users to access the data via a website for monitoring, remote control processing, and retrieval for statistical analysis. The temperature limits may change depending on certain factors of broiler management.

Our simulation proposal applies to tropical coastal areas at sea level, replicable to other environments based on specifications asked by broiler farmers. The purpose is to offer an IoT-based solution for broiler management through simulation with ICT tools to optimize automation processes for food distribution and temperature control to improve productivity by focusing on good practices, ensuring access to quality food for the population and responsible rural production, outcomes that align with the \footnote{https://www.undp.org/sustainable-development-goals}United Nations Sustainable Development Goals 2 (Rural Development, Food Security and Nutrition) and Goals 12 (Sustainable Consumption and Production).

The organization of this work has the following structure: Section \ref{Related Work} presents a short related work on systems that approach ICT for broiler production. Section \ref{Methods} focuses on the methods used for the simulation of the proposed network system, describing the materials used and the integration of the components. The results of the system tests and simulation scenarios are presented in Section \ref{Results}. Finally, Section \ref{Conclusions} presents the conclusions.

\section{Related Work}
\label{Related Work}

Studies focused on technical-level broiler management present different features and approaches. For instance, Elham et al. \cite{Elham2020} monitor the environment of poultry farms using IoT and Blockchain. The methodology consisted of acquiring information through IoT-based temperature sensors for data processing in JSON format using IOTA, an integrated technology between IoT and blockchain. This method allows computers on the IOTA network to convey data and immutable values to each other. Similarly, Watthanawisuth et al. \cite{Watthanawisuth2009} introduced ZigBee technology for a wireless interconnection of a sensor network with the advantage of operating at low energy consumption. In this case, the system monitors the air temperature microclimate with a high degree of coverage through a cluster tree topology to ensure the necessary scope using a few sensor nodes. The microcontroller interacts with duly protected temperature sensors by a cylindrical casing to avoid errors during the variables' or interference measurements. A photoelectric cell energizes the sensors through an energy storage circuit to allow the sensors to work even in the evenings, and a receptor processes the data in determined periods depending on the power required by each node.

Choukidar and N. Dawande \cite{Choukidar2017} employed sensors for level measurements of food, gas, water, and temperature to maintain environmental conditions within the farm. They connected the sensors to a Raspberry Pi for data monitoring and GPRS-based control to allow the farm administrators to consult the data gathered by the sensors through a web application. Following automation processes, they kept the parameters in the appropriate ranges by combining logical and physical elements. Their findings include an incidence reduction of environmental conditions that affect the weight and health of the poultry, leveraging an opportunity to save costs. Thus, the cases presented in this section approach to applying microcontrollers and microprocessors in interaction with sensors to convey information through wireless technologies to enhance farming processes.

\section{Methods}
\label{Methods}

\begin{figure}[htbp]
\centerline{\includegraphics[height=65mm]{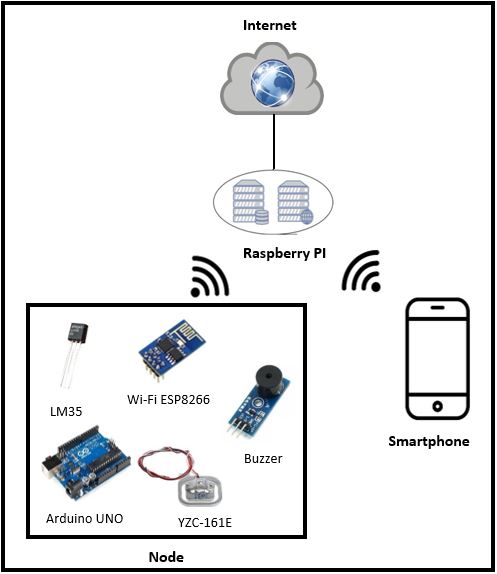}}
\caption{System topology.}
\label{topology}
\end{figure}

We implemented a single-node configuration using a temperature sensor without communication between more sensors because of the potential of the adaptability of radio modules for data conveying to establish transmission with microcontrollers and microprocessors. Further, photovoltaic cells supply energy for sensors to measure temperature and food loading and show information through end-user interfaces from the web application, such as Ubidots, allowing an interaction with our devices in a programmatic way through a friendly user interface with cloud-based services for IoT \cite{9526247, Enciso2018, 8549759}. The importance of these parameters lies in their direct influence on the breeding in which the information transmission is through a Wi-Fi module to attain lower budgets. A similarity of our work with related literature is the star topology-based deployment to achieve communication with a central node and data conveying. We considered this topology convenient for our network system because of its advantage of operating at low power consumption \cite{8985889}. We relied on the cloud for later dashboard visualization, as depicted in Fig. \ref{topology}.

\subsection{Materials}

We deemed the simulation scenario on the materials described because they are convenient for the conditions presented, representing low-cost alternatives for the purpose we followed. However, other approaches with different requirements can use similar software and hardware. The materials are listed as follows: 

\begin{figure}[htbp]
\centerline{\includegraphics[width=7cm, height=4cm]{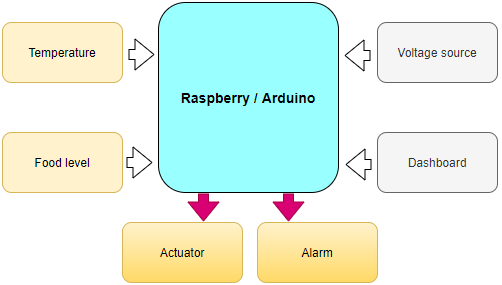}}
\caption{Block diagram.}
\label{blockdiagram}
\end{figure}

\begin{figure*}[!ht]
  \centering   
  \includegraphics[height=70mm]{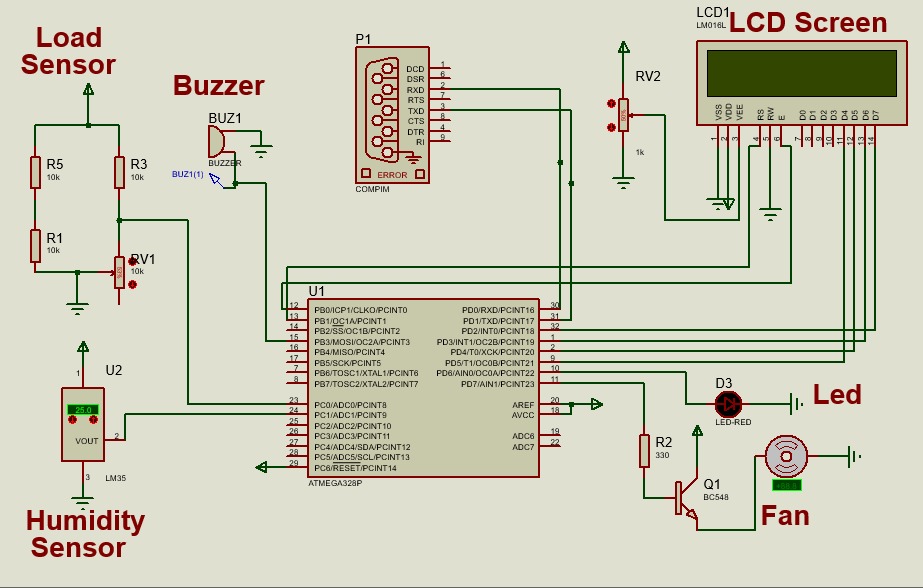}
  \caption{Proteus sensor node diagram.}
  \label{sensornodediagram}
\end{figure*}

\begin{itemize}
    \item {\footnote{https://www.ti.com/document-viewer/LM35/datasheet}LM35: Sensor that has a calibrated precision of 1 linear centigrade degree. It measures the temperature of the environment and returns a value in volts. The ratio of 1°C equals 10mV \cite{ramos2017characterization}.}
    
    \item {\footnote{https://datasheethub.com/yzc-131-10kg-range-weighing-sensor-load-cell-sensor/}YZC-161E: Sensor to measure either the load or weight of an object. It has a nominal load of 5 kg with an operating temperature range of -21 $ \sim$ 40°C. It is recommendable to work with an excitation voltage of 5 Vdc and a weight of 80 g.}
    
    \item {Arduino UNO: Microcontroller for running and controlling programmed activities via sensor interaction \cite{Chatterjee2017}. It comprises an ATmega328 chip, 5V operating voltage, direct current per I/O Pin of 40mA, 32 KB Flash memory, and 2 KB SRAM.}
    
    \item {ESP8266: A Wi-Fi module for wireless communication between modules through telemetry protocols \cite{Stoev2020, ESP8266}. It uses a Tensilica L106 32-bit CPU, 3V to 3.6V operating voltage, and 80 mA. It supports IPv4 and TCP/UDP/HTTP/FTP protocols.}
    
    \item {Raspberry PI 3 model B+: Electronic development board for information processing and analysis. It comprises a CPU + GPU with Broadcom BCM2837B0 card, 64 bits through 1.4GHz, with 1GB LPDDR2 SDRAM RAM. Its communication circuitry supports 2.4GHz and 5GHz bands, with IEEE 802.11 and Bluetooth protocols \cite{RaspberryPI3B+}}.
    
    \item {Buzzer: A module for alerting through sound with a 5Vdc passive buzzer, three pins, VCC +, GND -, and I/O signal \cite{Arduinomodules}. It requires a frequency of at least 1.5 kHz. Its advantage lies in its compact size.}
    
\end{itemize}

\subsection{Modular integration}

Fig. \ref{blockdiagram} shows a block diagram with the interaction of the functionalities of the components. The temperature and food level sensors convey the data obtained to the Raspberry/Arduino module for information processing. Based on the configured values, an alarm triggers to indicate the action to take, i.e., if the temperature values are very high, a mechanism activates to regulate the temperature. Likewise, the system will fill the reservoirs to supply food to the broiler if the food level in the dispenser is low. The data gathered are saved and displayed in a web application through a dashboard.

\begin{figure*}[!ht]
  \centering   
  \includegraphics[height=65mm]{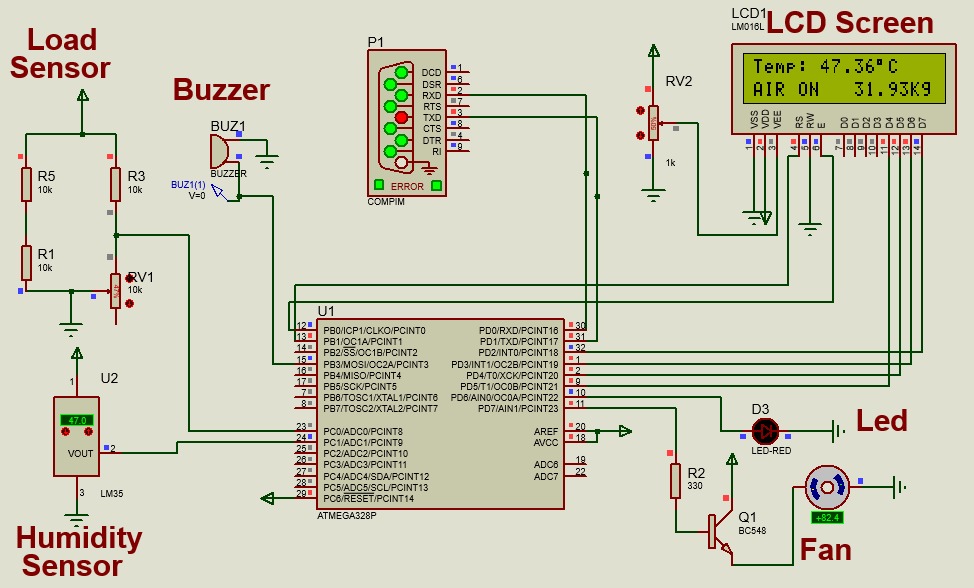}
  \caption{First scenario simulated}
  \label{simulation1}
\end{figure*}

\begin{figure*}[!ht]
  \centering   
  \includegraphics[height=65mm]{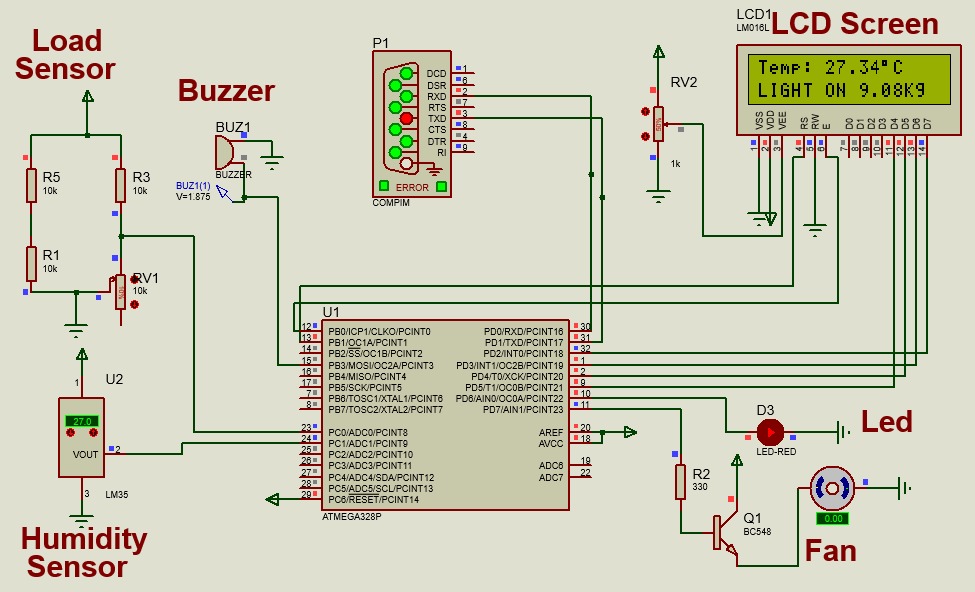}
  \caption{Second scenario simulated}
  \label{simulation3}
\end{figure*}

The sensor node integrated the temperature sensor, load sensor, buzzer module, Wi-Fi interface, and Arduino board. The LM35 sensor provides an output voltage that is proportional to the temperature being measured \cite{ramos2017characterization}, allowing us to measure the temperature integrating the Arduino, Raspberry PI, and Nodemcu IoT platform \cite{boonchieng2018smart}. The power control works through the YZC-161E Strain Gauge, setting it at 50 kg, while the strain gauge enables a transducer to support compression, tension or bending loads to convert force into an electrical signal.

The sensors were located at the bottom of the reservoirs to measure the food storage level, in which the reservoirs store the food with no more than 50 kg of weight, alerting the user if they weight less than 10 kg to indicate the food storage restocking. The LM35 and YZC-161E sensors communicated with the Arduino UNO board to assess if the temperature and food level parameters were correct. Otherwise, it sends the signal to the buzzer module to issue an audible alert for a re-calibration set by the administrator. The Raspberry Pi received the data through the ESP8266 Wi-Fi module to make possible the visualization of the web application to depict the variable status, i.e., the temperature and the food level to supply the broiler.

The integration of the sensor as a node allowed measuring of the temperature and food distribution controlled by the Arduino UNO, which saved the information for transmission to the Gateway node. It is possible to verify the functionality of the Gateway node at the web and database service level implemented in the Raspberry Pi \cite{Zainudin2019, Ortiz2018, Enriquez2015}. In our case, we chose to work on Raspbian Linux to setup the graphical interface for the IoT-based embedded system \cite{antony2020review}.

We used Proteus Design Suite software because it allowed electronic design automation. Fig. \ref{sensornodediagram} shows the schematic diagram of the temperature and food level monitoring system. The LM35 measured the temperature through a variation-based comparison, sending an analog signal convert it to a digital value. We have used an array of resistors to simulate the strain gauges. In one of them, we considered a potentiometer to vary the loading through an analog value to set the maximum and minimum values to obtain the results in kilograms.

\section{Results}
\label{Results}

\subsection{System testing}

The network system has visual and sound indications to optimize broiler management. We established an ideal temperature of 39°C. If the temperature got lower, the function turned on a light bulb. In the diagram, the LED represents this function, so if the temperature was higher than the ideal, a fan represented by a motor turned on. For the food loading in the dispenser, we established a minimum value of 10 Kg. If the sensor reads less than this value, the system emits a sound alert through a buzzer for a determined time.

The Proteus COMPIM simulated the sensor node connectivity with the Raspberry Pi through serial communication. We can remark that the LCD allowed visual indication of the data gathered by the sensors. In addition, Ubidots performed as a dashboard by device creation, and a token conveyed the data to this module. We set two variables called Load and Temperature, each with an ID used in the coding to know which variable transmitted the temperature and loading data.     

\begin{figure*}[!ht]
  \centering   
  \includegraphics[height=65mm]{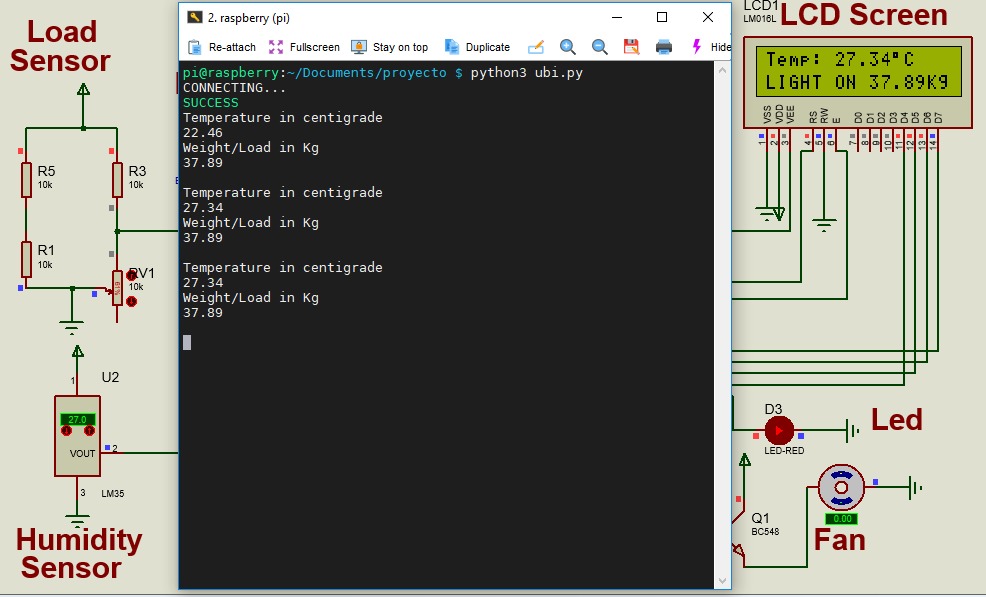}
  \caption{Data reception}
  \label{datareception}
\end{figure*}

\begin{figure*}[!ht]
  \centering   
  \includegraphics[height=50mm]{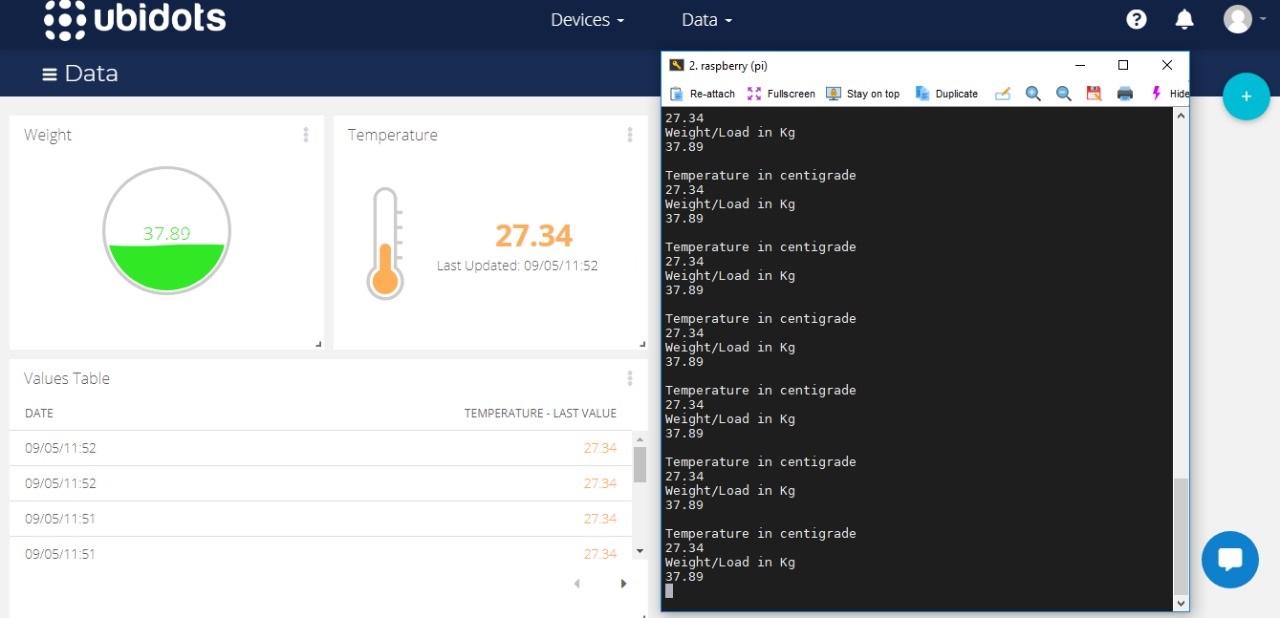}
  \caption{Dashboard}
  \label{dashboard}
\end{figure*}

\subsection{Scenarios simulated}

%\begin{figure*}[!ht]
%  \centering   
%  \includegraphics[height=82mm]{simulacion1.jpeg}
%  \caption{Scenario of temperature higher than allowed.}
%  \label{systemarchitecture}
%\end{figure*}

%\begin{figure*}[!ht]
%  \centering   
%  \includegraphics[height=82mm]{simulacion2.jpeg}
%  \caption{Low temperature scenario than allowed}
%  \label{systemarchitecture}
%\end{figure*}

%\begin{figure*}[!ht]
%  \centering   
%  \includegraphics[height=82mm]{simulacion3.jpeg}
%  \caption{Scenario of temperature lower than allowed and level of %food load lower than established.}
%  \label{systemarchitecture}
%\end{figure*}

The first scenario consisted of an environment at 47.36°C. Since this temperature was higher than 39°C (ideal room temperature), the LED remained off, and the motor (fan) turned on to decrease the room temperature. Fig. \ref{simulation1} depicts the fan speed in revolutions per minute (RPM) from the Proteus simulation. The buzzer did not turn on since the food amount level was above 10 kg. Fig. \ref{simulation3} presents a second scenario with a temperature of 27.34°C (less than 39°C). The LED did light up with the motor at 0 RPM. The buzzer remained off since it did not meet the conditions for turning it on.

%\begin{figure*}[htbp]
    %\centerline{\includegraphics[height=45mm]{Resultados1.jpeg}}
    %\caption{Temperature}
    %\label{temperatureresults}
%\end{figure*}

%\begin{figure*}[htbp]
    %\centerline{\includegraphics[height=45mm]{Resultados2.jpeg}}
    %\caption{Food load}
    %\label{foodloadresults}
%\end{figure*}

We reduced the potentiometer value for shifting the food loading level to 9.08 kg (less than 10 kg). It caused the buzzer to emit a 2-second interval sound, which remained in alert mode until loading the food dispenser again. We configured a Python script that received the sensor data through a serial connection with Proteus on a virtual machine with Raspbian. The script transmitted the data to Ubidots through the API client for a dashboard visualization. Fig. \ref{datareception} shows a transition between states, such as connecting and success, representing the connection between Proteus and Ubidots. The temperature and loading values were in kilograms and displayed through the console. The system provided real-time updates to check the food loading and temperature through the dashboard, as shown in Fig. \ref{dashboard}.

\begin{figure*}[!t]
\centering
\subfloat[Temperature]{\includegraphics[height=40mm]{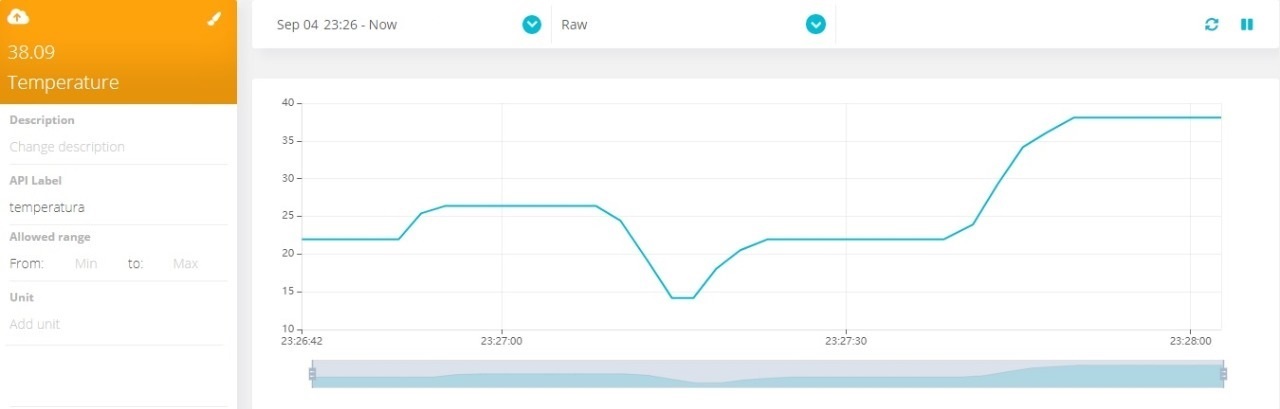}%
\label{temperatureresults}}
\hfil
\subfloat[Food load]{\includegraphics[height=40mm]{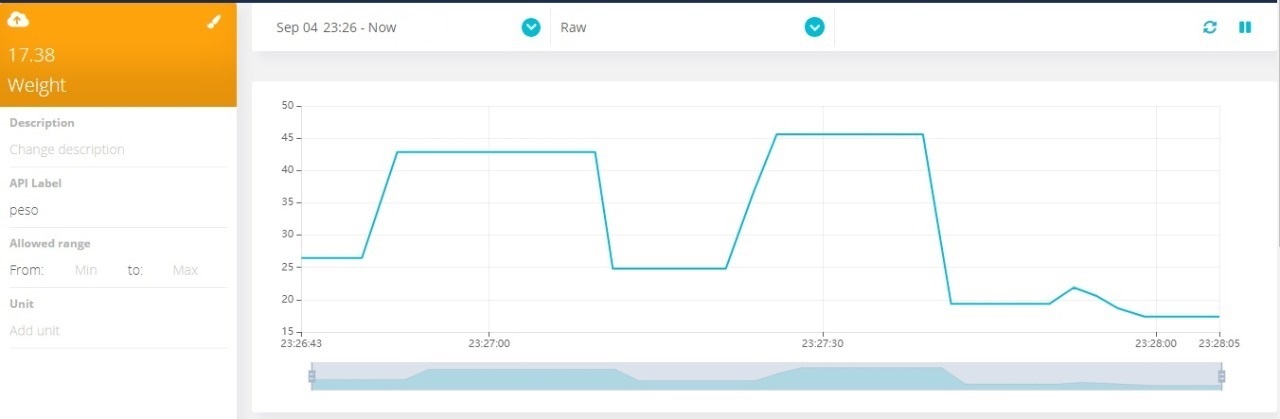}%
\label{foodloadresults}}
\caption{Screenshot of the results}
\label{fig:resultsTempFood}
\end{figure*}

\subsection{Analysis}

\begin{figure}[htbp]
    \centerline{\includegraphics[width=7.5cm, height=2cm]{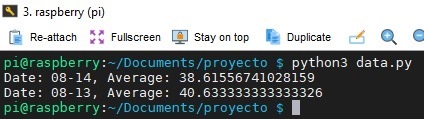}}
    \caption{Two-day monitoring mean (average)}
    \label{resultanalysis}
\end{figure}

This work simulated a broiler farm environment based on experiments. Therefore, the data gathered can resemble contexts where the temperature and load sensors undergo similar conditions. Fig. \ref{fig:resultsTempFood} represents the temperature obtained in two days (\ref{temperatureresults}) and the values of food load in the same period (\ref{foodloadresults}). It is possible to export the results gathered from Ubidots during the simulation into a \emph{.csv} format and perform data filtering for specific intervals. A script allows manipulating the data collected in the \emph{.csv} file with 1103 values on different schedules. The sum of data from a day was gathered and divided by the amount of data to obtain the mean for two consecutive days. On the first day, the mean was approximately 38.61°C and 40.63°C on the second day, as depicted in Fig. \ref{resultanalysis}. \textit{Appendix A} lists the Raspberry coding to integrate the results collected from Ubidots with the API for connection testing and to print the temperature and food load results.

%\begin{figure}[htbp]
%\centerline{\includegraphics[width=8cm, height=10cm]{Diagrama Esquematico1.JPG}}
%\caption{Schematic Diagram.}
%\label{schematicdiagram}
%\end{figure}

On both days, we expected a mean of 39°C to represent adequate conditions for broiler management. However, since the data taken to obtain the displayed mean comes from the direct manipulation of people, we do not recommend a comparison with an average taken from field implementations. We verify that our work meets the proposed objectives, such as issuing visual and sound alerts and capturing changes in the variables of primary interest. Thus, we contributed to resource optimization and ensured conditions for recommendable broiler management.

\section{Conclusion}
\label{Conclusions}

The simulation yielded accurate results for the proposed objectives, as the temperature and load variables responded effectively to the external factors detected by the system. However, this proposal must also adhere to the regulations and policies of the implementation site. Depending on the context, the integration of robust sensors can ensure reliable data acquisition. This does not necessarily lead to higher equipment expenditures, but may require investment in training for managers and staff. It is important for the readers to consider that higher-capacity electronic development boards (e.g., FPGAs) can enhance implementation dynamics, particularly when it comes to sensor integration. These boards align with wireless communication concepts, supporting components that transmit through different protocols and channel frequencies.

In addition, a reliable circuit power supply must always be in place to handle power failure events, incorporating backups such as a battery bank or uninterrupted power sources. Actuator and alert element tests should include a backup system, such as SMS warning logs, to verify the information displayed on the dashboard, either during a failure or inaccessibility. For future work, we recommend integrating other open-source technologies to address budget concerns, especially if a deployment based on our guidelines needs to expand its scope and follow specific rules.

\ifCLASSOPTIONcaptionsoff
  \newpage
\fi

\appendices

{\section*{\\Appendix A}

\begin{lstlisting}[language=Python, caption=Coding for test connectivity and printing results]
from ubidots import ApiClient
import serial
import time

if __name__=='__main__':
    dato=0
    try:
        print("Connecting...")
        arduino=serial.Serial('/dev/ttyS3', 9600)
        time.sleep(1)
    except:
        print("Fail to connect")
    try:
        print("Connecting API...")
        api=ApiClient(token='tokenID')
        temperature=api.get_variable('ID1')
        weight=api.get_variable('ID2')
        print('testing connectivity')
    except:
        print("Fail to connect API")

    while True:
        temp=arduino.readline()
        dato=float(temp)
        print("Temperature in Centigrade")
        print(dato)
        newTemp=temperature.save_value({'value':dato})
        load=arduino.readline()
        c=float(load)
        print("Weight/Load in Kg")
        print(c)
        newLoad=weight.save_value({'value':c})
        print("  ")

\end{lstlisting}

%\begin{IEEEbiography}[{\includegraphics[width=1in,height=1.25in,clip,keepaspectratio]{figure/unknowngirl.png}}]{Sandra Isabella Coello}
%\lipsum[1-2] holds a B.Sc. in Telematics Engineering from the Escuela Superior Politécnica del Litoral, Guayaquil, Ecuador.
%\end{IEEEbiography}

%\begin{IEEEbiography}[{
%\includegraphics[width=1in,height=1.25in,clip,keepaspectratio]{figure/unknowngirl.png}}]
%{V. Sanchez Padilla}
%(Member, IEEE), received a master’s degree in telecommunications engineering from George Mason University, Fairfax, VA, USA, and a master’s degree in productivity and quality management from the Escuela Superior Politécnica del Litoral, Guayaquil, Ecuador. 
%\end{IEEEbiography}

%\newpage

%\begin{IEEEbiography}[{\includegraphics[width=1in,height=1.25in,clip,keepaspectratio]{figure/unknownman.png}}]{Ronald Ponguillo-Intriago} received a master's degree in management information systems from the Escuela Superior Politécnica del Litoral, Guayaquil, Ecuador. 
%\lipsum[1]
%\end{IEEEbiography}

%\begin{IEEEbiography}[{\includegraphics[width=1in,height=1.25in,clip,keepaspectratio]{figure/unknowngirl.png}}]{Maya Menon}
%\lipsum[1-2] holds a Ph.D. in engineering education from Virginia Polytechnic Institute and State University, a M.Tech. in robotics and automation from Amrita University, and a B.S. in Computer Science from Arizona State University.
%\end{IEEEbiography}

\vspace{11cm}

%\printacronyms 
\end{document}